\documentclass{article}
\usepackage[utf8]{inputenc}

\usepackage[english]{babel}
\usepackage[T1]{fontenc}

\usepackage[a4paper,top=3cm,bottom=2cm,left=3cm,right=3cm,marginparwidth=1.75cm]{geometry}

\usepackage{amsmath}
\usepackage{graphicx}
\usepackage[colorinlistoftodos]{todonotes}
\usepackage[colorlinks=true, allcolors=blue]{hyperref}
\title{Insider threat modeling: An adversarial risk analysis approach}
\author{Chaitanya Joshi, Jesus Rios and David Rios}
\date{ }
\begin{document}
\maketitle


\begin{abstract}
Insider threats entail major security issues in geopolitics, cyber risk management  and business
organization. The game theoretic models proposed so far do not take into account some important factors such as the organisational culture and whether the attacker was detected or not. They also fail to model the defensive mechanisms already put in place by an organisation to mitigate an insider attack. We propose two new models which incorporate these settings and hence are more realistic. 
We use the adversarial risk analysis (ARA) approach to find the solution to our models. ARA does not assume common knowledge and solves the problem from the point of view of one of the players, taking into account their knowledge and uncertainties regarding the choices available to them, to their adversaries, the possible outcomes, their utilities and their opponents' utilities. Our models and the ARA solutions are general and can be applied to most insider threat scenarios. A data security example illustrates the discussion.
\end{abstract}


\section{Modeling Insider Threat}


Insider threats are encountered in areas such as international security, geo-politics, business, trade and cyber security.  They are widely perceived to be  
significant (\cite{ca2018}, \cite{haystax2017}), and even often considered
to be more
damaging and likely than outsider attacks (\cite{ca2018}, \cite{cert2012}).
Moreover, it is feared that the impact of insider threats actually
known is only the 
tip of an iceberg, as many organizations are choosing not to report such
incidents unless required to do so
by law (\cite{sym2016}): as described in \cite{HUNKER}, it is a problem area 
 in which little data is available, specially in the cyber security domain.
Protection from insider threats is challenging as the perpetrators might have
access to  sensitive resources and privileged system accounts.
Solutions to insider threat problems are considered to be complex (\cite{DAT2015}): 
technical solutions do not suffice since these threats are fundamentally a people issue, as thoroughly discussed in \cite{SARKAR} and \cite{GREIZER}.

In its simplest form, it is natural to view the insider threat problem as a two player game. We may call the first player the {\em defending organization} or the {\em defender}
(which could refer to a single business or military unit or a similar entity, but also to a whole country or a coalition of entities or countries)
and the second one, the {\em malicious insider} or {\em attacker}
(which could refer to one or more employees, contractors, or persons who have significant 
access to the organization and have been trusted with such access). A typical scenario would be as follows:  
since insider threats are a well-known phenomena, it will frequently be the case
that several measures 
would have already been implemented by the organization (at least, 
if it is sufficiently mature) to prevent or deter insider attacks. 
As an example, \cite{SILOWASH} provides  a catalogue of best practices  against insider threats in the cybersecurity domain.
The insider will typically be aware of the measures in place and plans an attack accordingly. Once the attack has
been carried out and detected, the organization will undertake actions to end the attack and mitigate any damage caused, possibly based 
on the resources deployed at the first stage. This type of interactions have been named sequential
Defend-Attack-Defend games, e.g.\ in \cite{BROWN}. 

It is therefore natural that game-theoretic models of insider threats 
have been explored.  For example, \cite{liu08} model the problem as a two-player, zero-sum dynamic game. At each discrete time point, both players make decisions resulting in a change of state and opposite (given the zero-sum property) rewards to them. The authors then look for  \emph{Nash equilibria} (NE). While they model a stochastic game,
they assume that both the defender and the attacker have complete knowledge of each other's beliefs and preferences, which is hardly the case in applications.
Moreover, this model is oversimplified in several respects. For example, there could be multiple attackers, the attacker pay-offs might not be immediate to obtain and the game might not be zero-sum.  Also, in most cases, the defender would have already  employed measures to prevent an insider attack and, therefore, the problem  should be modeled  as a sequential Defend-Attack-Defend game, rather than as a simultaneous one.

A more realistic approach is described in \cite{kant10} who consider an insider threat problem in cybersecurity, trying to model the continuous interactions between an intruder and an intrusion detection system (IDS). They assume bounded rationality  on the agents, assess outcomes through utilities and 
use quantal response equilibria instead of standard NE. However, their model focuses on a particular application and is not immediately generalizable. Moreover, the game does not consider multiple          players and carries on even after detecting an attack. This is because they assume that detection causes the attack to be stopped, but does not eliminate the attacker from the game. \cite{tang11} also model insider threats to  IT systems considering bounded rationality, combining game theory with
an information fusion algorithm to improve upon traditional IDS based  methods by being able to consider various types of information. To combat Advanced Persistent Threats (APT) coupled with insider threats,
\cite{feng15} and \cite{hu15} propose three (defender, APT attacker and insider)
player games. They employ a two layer game and show the existence of NE.

Note that, non-game theoretic approaches have also been used to model insider threats. See, for example, \cite{martinez08} who employ a system dynamic approach, \cite{oliver12} who propose an approach that combines structural anomaly detection and psychological profiling and \cite{Axelrad13} who use a Bayesian network approach using a structural equations model. However, the focus of this paper is on game theoretic modeling and for that reason we do not consider non-game theoretic approaches in further detail.


In an insider attack, sometimes it may be difficult to detect who the attacker is since they already have access to the organisation, its premises, IT systems, etc. The malicious insider may be motivated to remain undetected not only to avoid punishment, loss of reputation, etc., but also to continue to keep causing harm or gain benefit, whatever the case may be. In the wake of the attack, an organisation would typically take actions, including improving their processes and defensive systems so that a similar attack is unlikely to succeed again in the future. However, in some cases, if the attacker has not been detected, then it may be very difficult to prevent a future attack regardless of any additional measures put in place. As an example, suppose that a malicious insider has shared sensitive information with a competitor and while the passing of 
information has been detected, the investigation has failed to identify the attacker. In this case, although the organisation could try and make it difficult for such information to be shared externally in the future, it may be quite difficult or impossible to guarantee that such sharing will not actually happen. Attacker detection is therefore an important consideration in the defender decision making 
and should be taken into account in a model. However, none of the game theoretic models proposed so far have taken this information into account.

The presence of a security culture has been identified as an important factor in preventing insider threat (\cite{Choi18}, \cite{Safa17}, \cite{SARKAR}). Moreover, it is widely accepted that having an unfavourable organisational culture will lead employees to override security policies and processes increasing the chances of insider attacks being successful (\cite{Probst10}, \cite{SARKAR}, \cite{Colwill09}). Further, the right culture could enhance levels of underlying personal trust, loyalty and mutual dependency (\cite{Colwill09}) thus reducing the chances of an insider attack being launched and also of it being successful if launched.   Yet the game theoretic models for insider threat proposed so far do not
either take into account the culture in the organisation.

An insider threat problem could be a multi-player game with possibly, multiple attackers and defenders. This could either be because of a group of attackers working together to harm the organisation or attackers acting independently and without the knowledge of each other,  
and, similarly, for the defenders.  \cite{liu08} acknowledge that we typically 
face a multi-player
game with multiple attackers and defenders in this domain.
However, because that is hard to solve, they club all attackers as a single player and assume that all defenders are coordinated, thus simplifying to a 2-player game. This could be a reasonable modeling approach since if the attackers are working in a team, then they could be considered to effectively conform 
a single entity and if they are independent individual attackers, 
each of those attacks could be modeled separately. However, an insider threat could also be a multi-player game due to the presence of different types of attackers. It is well known that not all attackers are malicious, many are inadvertent. Some of the previous research has focused on segmenting the employees. For example,  \cite{liu*} provide a segmentation with inadvertent and malicious insiders. However, the exact segmentation (good employees, inadvertent, malicious and so on) could vary from organisation to organisation. Also, the same employees are also the defenders and each employee will be differently effective as a defender based on the type of the employee and also the role they play inside the organisation. Finally, the exact mix of employees that a malicious insider might come across and how important a role each of them play in mitigating a malicious attack will vary in every situation. Therefore, it is impossible to model every possible scenario using a segmentation approach or indeed a multi-player approach. 

  Insider threat is a well known phenomenon and it is reasonable to expect that an organisation would have certain defensive measures in place to prevent an insider threat. Yet, the models proposed so far do not take into account the existing defensive measures or the effects they may have in either the prevention or the detection of an  attack. For example, \cite{kant10} propose a sequential game in which the attacker performs first. They do not acknowledge the existence of defensive processes in place to prevent 
  insider attacks. However, modeling the initial defensive measures is important since it is well known (\cite{moore15}, \cite{liu*}, \cite{martinez08}) that such measures could impact the organisational culture including unintended negative consequences. While such measures may improve physical security or the security culture (as intended), they could also make employees feel not trusted or being intruded upon. The latter case may lead to a culture of mistrust and may cause employees to not follow the processes or find ways to get around them thus increasing the chances of an insider attack success.  Ours is the first paper to model consider the initial defensive mechanism and model the insider threat as a defence-attack-defence game.
 

Indeed, this paper makes two main contributions. Firstly, it proposes two novel, more realistic, defend-attack-defend models for the general insider threat problem: one accounts for whether the attacker has been detected or not;  the second one, that, in addition, also takes into account the organisation culture, even accommodating the possibility that the attack was prevented altogether. Secondly, we solve these models using the Adversarial Risk Analysis (ARA) approach thus ensuring that the models do not make any unrealistic common knowledge assumptions, only take into consideration the information that is likely to be available to the defender and are practicably solvable.
When modeling the insider threat, our focus is on modeling the
agents' interactions and not on the (anomaly) detection or any other aspect. Specifically, we focus on finding the optimal initial defensive mechanism that the defender should install to mitigate the insider threat and the optimal defensive action that the defender should take in the wake of an insider attack given the initial defensive action, the attack and its outcome.

%

The structure of the paper is as follows. In Section \ref{ARA} we provide a brief overview of ARA, show why a solution almost always exists and illustrate how to find Monte Carlo solutions to the problems. Next, we propose our first new model and show how it can be solved using ARA. We propose our second new model in Section \ref{realistic} and show how an ARA solution for that problem can be found. In Section \ref{exe}, we use a numerical example to illustrate how both of the proposed models could be used to
deal with an insider threat problem. Finally, in Section \ref{discuss} we summarise and discuss the challenges in implementing ARA and briefly highlight the scope for further work.\\

\section{Adversarial Risk Analysis} \label{ARA}

While game theory has been the typical choice to model interactions between two or
more strategic adversaries, limitations of such theory
have been pointed out, focusing on the common knowledge assumption
and the conservative nature of its solutions,  e.g.\ \cite{gintis09}, \cite{camerer03},
or \cite{raiffa02}.  Bayesian game theory (\cite{harsanyi67}) can be used to model games with imperfect information by eliciting prior distributions on different \emph{types} of opponents. However, this approach requires that the prior distributions elicited by each player be commonly known. This assumption is unrealistic and while methods have been proposed to implement Bayesian game theory without making the common prior assumption (for example, \cite{Antos10},  \cite{Sakovics01}), these have not caught on. The main challenge in game theory is that it aims to find solution for all the players and moreover this solution needs to be an equilibrium solution. This makes it increasingly difficult to find solutions as the models get more realistic and complex.

On the other hand, while conventional risk analysis does not assume common knowledge and solves the problem only for one of the players, it cannot model the strategic thinking of an intelligent adversary. Limitations of conventional risk analysis in security have been pointed out as well, \cite{cox09a} and \cite{brown11}.

Adversarial risk analysis (ARA) (\cite{rios09}) was proposed to address the shortcomings of both the game theory and risk analysis approaches. ARA does not assume common knowledge and  solves the problem from the point of view of just one of the players, taking into account their knowledge and uncertainties regarding the choices available to them, to their adversaries, the possible outcomes, their utilities and their opponents' utilities. ARA takes into account the expected utilities for the defender as well as the random expected utilities for the opponents, incorporating uncertainty regarding their
strategic reasoning.
Since its introduction, it has been used to model a variety of problems such as network routing for insurgency (\cite{wang11}), international  piracy (\cite{sevillano12}), counter-terrorism (\cite{rios12}),  autonomous social agents (\cite{esteban14}, urban security resource allocation (\cite{gil16}), 
adversarial classification (\cite{naviero19}), counter-terrorist online surveillance (\cite{Gil19}), cyber-security (\cite{rios19}). 

One of the distinguishing aspects of ARA compared to game theory is that, ARA only solves the problem for one of the players (typically, the defender). Unlike game theory, it does not aim to find an equilibrium solution for all the players. Since ARA aims to find a solution that is optimal only for one of the players, solving ARA is relatively easy. In fact, as we show below, solving ARA essentially involves taking expectations (using MC integration) and finding the maximum over a low-dimensional space (often these spaces are discrete, at least on some of the dimensions) and therefore, a solution almost always exists.\\
Further, unlike game theory, ARA only considers the information available to the player and their uncertainties when solving the problem. It does not have to consider the information available to each player about the rest of the players. Thus for a basic $n$- player simultaneous game, ARA would only require eliciting $(n-1)$ probability distributions, but (Bayesian) game theory would require eliciting  $n \times (n-1)$ probability distributions (\cite{liu08}). 

\begin{figure} [!h] 
{\centering
\includegraphics [scale=0.45]{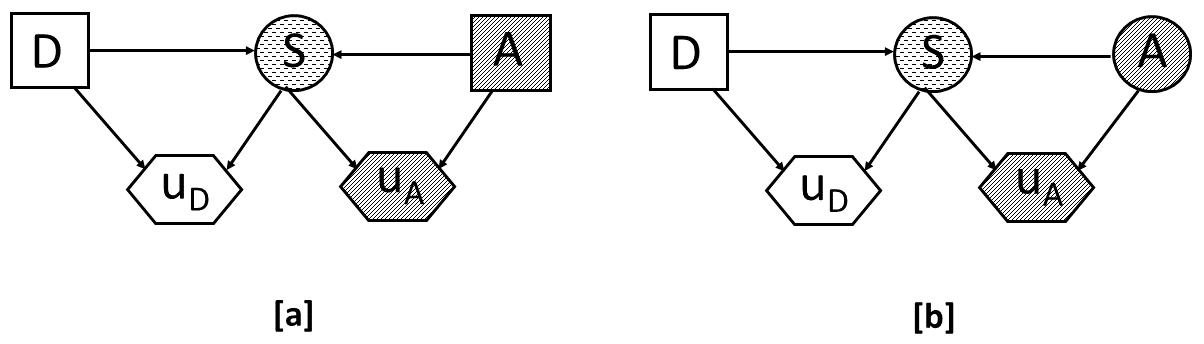}
\caption{[a] BAID for a two player simultaneous game and [b] BAID from the point of view of just the defender $D$}
\label{BAID_fig}
}
\end{figure}

To illustrate how a basic game can be modeled using ARA, we consider a two player simultaneous game between a defender $D$ (she) and an attacker $A$ (he). Figure \ref{BAID_fig}[a] presents the problem using a bi-agent influence diagram (BAID) where decisions are represented by square nodes, uncertainties with circular nodes and utilities  with hexagonal nodes. Nodes corresponding solely to $D$ are not shaded; 
those corresponding exclusively to $A$ are diagonally shaded; finally, shared chance
nodes are shaded using horizontal dashed lines. Suppose we are solving the problem for the defender $D,$ Figure \ref{id_1}[b] represents the BAID from her point of view. The only difference here is that $A$ is also a random node now since $D$ is not certain as to what action $A$ would take. Let $\mathcal{D}, \mathcal{A} \mbox{ and } \mathcal{S}$ denote the set of all possible actions for the defender, the set of all possible actions for the attacker and the set of all possible outcomes, respectively.

The basic idea behind finding a standard ARA solution is to \emph{integrate out} the uncertainties at the random nodes and \emph{maximize} the expected utility at the decision nodes. In this problem, $D$ has to consider two random nodes, $A$ and $S.$ ARA 
can be solved using backward induction: we start from the final (random/decision) node and move backwards. The defender will try to find her best action as follows.

First, the defender needs to find her expected utility by taking into account the uncertainty due to the random outcome $s$ as
\begin{equation} \label{Bi-1}
     \psi_{D}(d,a) = \sum_{s\in\mathcal{S}} u_{D}(d,a,s) p_{D}(s|d,a)\, \mbox{ for all } d \in \mathcal{D}, a\in\mathcal{A}  .
\end{equation}
Next, she needs to find her expected utility by considering the uncertainty in the random action node $A$ as
\begin{equation} \label{Bi-2}
    \psi_{D}(d) = \sum_{a\in\mathcal{A}} \psi_{D}(d,a) p_{D}(a).
\end{equation}
Finally, she can find her optimal action $d^{*}$ as 
\begin{equation} \label{Bi-3}
    d^{*} = \arg\max_{d\in\mathcal{D}} \psi_{D}(d).
\end{equation}
This requires the defender to elicit the uncertainties in the random nodes using probability distributions $p_{D}(a)$ and $p_{D}(s|d,a)$. Of these, $p_{D}(s|d,a)$ can typically be elicited based on her knowledge/experience and/or her subjective beliefs. She can choose to elicit $p_{D}(a)$ also using her beliefs, knowledge or past experience. However, she can also choose elicit it by modeling the strategic thinking of her adversary. For example, she may believe that like her, $A$ is also an expected utility maximizer, who would choose the action $a^{*}$ maximizing his expected utility. She can determine $a^{*}$ by solving  Equations (\ref{Bi-1}) - (\ref{Bi-3}) for the attacker using his utility function $u_{A}(d,a,s)$ and his probabilities $p_{A}(d)$ and $p_{A}(s|d,a)$ instead. However, these will typically be unavailable to her. She can account for her uncertainty in the utility and the probabilities of the attacker, eliciting random utility and random probabilities $U_{A}(d,a,s), P_{A}(d)$ and $P_{A}(s|d,a).$ She can then find the random expected utility for the attacker through 
\begin{equation} \label{Bi-4}
    \Psi_{A}(a) = \sum_{d} \left ( \sum_{s} U_{A}(d,a,s) P_{A}(s|d,a)  \right ) P_{A}(d) \,.
\end{equation}
Then, she can find his random optimal action $A^{*}$ as
\begin{equation} \label{Bi-5}
    A^{*} = \arg\max_{a\in\mathcal{A}} \Psi_{A}(a).
\end{equation}
Finally, once the defender assesses $A^{*},$ her predictive distribution about the attack's action, she is able to solve her decision problem.
We have assumed in our notation above that the sets of actions $\mathcal{D}$ and $\mathcal{A}$ for the defender and the attacker respectively as well as the set of outcomes $\mathcal{S}$ are discrete. If any of these sets were continues we would have integrals instead of summations.

If computing expectations using integrals over probability density functions in Equations (\ref{Bi-1}) and (\ref{Bi-2}) or maximizing a univariate function in Equation (\ref{Bi-3}) cannot be performed analytically, we can always approximate them using Monte Carlo methods.\\

\begin{underline}{Monte Carlo algorithm to solve the defender's problem} \end{underline}
\begin{enumerate}
    \item For each $d$ in a grid $\{d_1,\ldots,d_n\} \subset \mathcal{D}$:
    \begin{enumerate}
        \item For $k=1,\ldots, N,$ 
        \item[] $\quad$ sample $a_{k}\sim p_{D}(a)$ 
        \item[] $\quad$ sample $s_{k}\sim p_{D}(s|d,a_k)$
        \item[] $\quad$ compute $\psi_{D}^k(d) = u_{D}(d,a_k,s_k) $
        \item $\hat{\psi}_{D}(d) = \frac{1}{N} \sum_{k=1}^{N} \psi_{D}^{k}(d)$
     \end{enumerate}
    \item Find $d^{*} =\arg\max \hat{\psi}_{D}(d_{i}), 1\leq i \leq n \, .$
\end{enumerate} 


When $p_{D}(a)$ is elicited using Equations (\ref{Bi-4}) and (\ref{Bi-5}), 
then it can also be approximated using Monte Carlo methods. For this, one must elicit the probability distributions for the random probabilities $P_{A}(d)$,  $P_{A}(s|d,a)$, and random utility $U_{A}(d,a,s)$. 

\pagebreak

\begin{underline}{Monte Carlo algorithm to elicit $p_{D}(a)$} \end{underline}
\begin{enumerate}
    \item For $k=1,\ldots N$:
    \begin{enumerate}
        \item Sample attacker's probabilities and utility
        \item[] $\quad$ $u_{A}^{k}(d,a,s) \sim U_{A}(d,a,s)$
        \item[] $\quad$ $p_{A}^{k}(s|d,a) \sim P_{A}(s|d,a)$
        \item[] $\quad$ $p_{A}^{k}(d) \sim P_{A}(d)$
        \item Calculate $\psi_{A}^{k}(a) = \sum_d \left ( \sum_s u_{A}^{k}(d,a,s) p_{A}^{k}(s|d,a)\right ) p_{A}^{k}(d) \sim \Psi_{A}(a) .$
        \item Find $a^{*}_k = \arg\max \psi_{A}^{k}(a) \sim A^*$
    \end{enumerate}
    \item If $\mathcal{A}$ is discrete  
        \item[] $\quad$ approximate $p_{D}(a) = Pr( A^{*} = a ) $ by $ \frac{1}{N} \sum_{k=1}^{N} 1(a^{*}_k  = a   ),$ for all $a \in \mathcal{A}$ ,
    \item[] else, if  $\mathcal{A}$ is continuous 
        \item[] $\quad$ approximate $p_{D}(a) = Pr( A^{*} \leq a ) $ by $ \frac{1}{N} \sum_{k=1}^{N} 1(a^{*}_k  \leq a   ),$ for any give $a \in \mathcal{A}$ .
\end{enumerate}

We have assumed that the attacker is an expected utility maximizer to assess $p_{D}(a)$. However, this probability can also be assessed assuming the attacker uses a number of alternative solution concepts other than expected utility maximization. For example, the defender could solve the problem by assuming that the attacker is a non-strategic player or that he is level-$k$ thinker or uses NE to determine his optimal action and so on. An ARA solution can be found for each of these solution concepts, see \cite{banks16} for details.\\

\section{An ARA Defend-Attack-Defend model} \label{kaka} 

\begin{figure} [!h] 
{\centering
\includegraphics [scale=0.45]{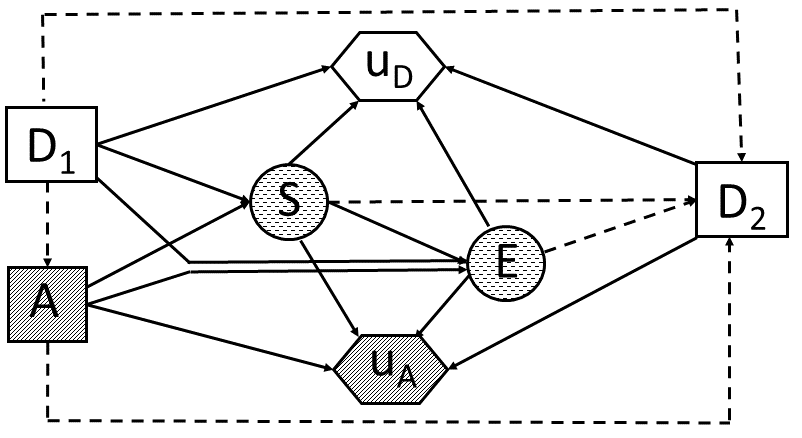}
\caption{BAID for the Defend-Attack-Defend  insider threat game}
\label{id_1}
}
\end{figure}

We start with a Defend-Attack-Defend model to deal with the insider threat problem, which considers a defender $D$ (the organisation, she) and an attacker $A$ (the insider, he). Our model is based upon the graphical framework described in \cite{banks16}. Figure \ref{id_1} presents the problem using a bi-agent influence diagram (BAID) where decisions are represented by square nodes, uncertainties using circular nodes and utilities  with hexagonal nodes. Nodes corresponding solely to $D$ are not shaded; 
those corresponding exclusively to $A$ are diagonally shaded; and, finally, shared chance
nodes are shaded using horizontal dashed lines. Dashed arrows indicate that the involved decisions are made with the corresponding agent knowing the values of the preceding  nodes, whereas solid arrows indicate probabilistic  or value dependence of the corresponding node with respect to its predecessors.

The action and outcome sets are as follows. Initially, the defender 
must choose one of the portfolios of preventive measures $d_1$ in set  $\mathcal{D}_1 $.
 Having observed the  portfolio implemented, the insider 
  will adopt  one of the actions $a$ in $\mathcal{A}$; this set
 could consist of either 'no attack' or 'attack' decisions or different types/intensities of attacks or other  attack options.  The set $\mathcal{S}$ consists of  the possible outcomes $s$ that can occur as a result of the  preventive portfolio $d_{1}$ and the attack $a$ undertaken.
 It is possible that the identity of the attacker remains undetected even after observing $s$. Node $E$ represents the event of whether the attacker was detected or not. Once $E$ has been observed,
 the  organization will choose to carry out one of the actions $d_2$ in 
 $\mathcal{D}_2 $  to end the attack, limit any damage and  possibly pre-empt future attacks leading to the final outcomes
 of both agents, respectively evaluated through  their utility functions $u_D$ and $u_A$.  Note that all three sets $\mathcal{D}_1$, $\mathcal{A}$ and $\mathcal{D}_2$
 could contain a  {\em do nothing} action.

For its solution, the defender must first quantify the following:
\begin{enumerate}
\item The distribution $p_{D}(a|d_{1})$ modeling her beliefs about the attack $a$  chosen at node $A$ by the employee, given the chosen defense $d_{1}$.  
\item The distribution $p_{D}(s|d_{1},a)$ modeling her beliefs about the outcome $s$ of the attack, given $a$ and $d_{1}$.
\item The distribution $p_{D}(e|d_{1},a,s)$ modeling her beliefs about whether or not the attacker is detected  given $d_{1}$, $a$ and $s$.                     
\item Her utility function $u_{D}(d_{1},s,e,d_{2})$ which evaluates  the consequences associated with their first ($d_1$) and second ($d_2$) defensive actions as well as the outcome $s$ and whether the attacker was detected or not. It also includes any future consequences of the action $d_{2}$ as assessed by the defender when choosing $d_{2}.$
\end{enumerate}

\noindent Given these assessments, the defender first seeks to
find the action $d_{2}^{*}(d_{1},s,e)$ maximizing her utility 
\begin{equation}\label{sim1}
d_{2}^{*}(d_{1},s,e) = \arg\max_{d_{2}\in \mathcal{D}_{2}} u_{D}(d_{1},s,e,d_{2}), 
\end{equation}
leading to the best second defense when the first one was $d_1$, the outcome  was  $s$ and the detection result, $e$. She
seeks to compute the expected utility $\psi_{D}(d_{1},a)$ for
each $(d_{1},a)$ 
taking into account the uncertainty in 
$e$ 
and $s$ 
$s$ 
is defined through
\begin{equation}\label{sim2}
\psi_{D}(d_{1},a) = \sum_s \left (\sum_{e} u_{D}(d_{1},s,e,d_{2}^{*}(d_{1},s,e)) \, p_{D}(e|d_{1},a,s) \right ) \, p_{D}(s|d_{1},a) .
\end{equation}

\noindent Moving backwards, she computes her expected utility for each $d_{1} \in \mathcal{D}_{1}$ using the predictive distribution $p_{D}(a|d_{1})$  through 
\begin{equation}\label{sim3}
\psi_{D}(d_{1}) = \sum_a \psi_{D}(d_{1},a)p_{D}(a|d_{1}).
\end{equation}
\noindent Finally, the defender finds her maximum  expected utility decision $d_{1}^{*} = $ $\arg\max_{d_{1}\in {\cal D}_1}  \psi_{D}(d_{1}).$ This backward induction shows that the defender's optimal strategy is to first choose $d_{1}^{*}$ and, then, after having observed $s$ and $e$, choose $d_{2}^{*}(d_{1}^{*},s,e)$.

The above analysis requires the defender to elicit $p_{D} (a|d_{1})$. This can 
be done using risk analysis based approaches as in \cite{ezell10} or by modeling the strategic analysis process of 
the insider. For this, the defender should model the insider's strategic analysis by assuming 
that he will perform an analysis similar to hers to find his optimal attack $a^{*}$.
To do so, she should assess his utility function $u_{A}(a,s,e,d_{2})$ and probability 
distributions $p_{A}(e|d_{1},a,s),\, p_{A}(s|d_{1},a)$ and $p_{A}(d_{2}|d_{1},a,s,e).$ 
However, since the corresponding judgments will not be available to the defender, we could model her uncertainty about them through a random utility function $U_{A}(a,s,e,d_{2})$ and random probability distributions $P_{A}(e|d_{1},a,s), \, P_{A}(s|d_{1},a)$ and $P_{A}(d_{2}|d_{1},a,s,e).$ 
There are multiple ways for the defender to elicit these random utilities and probabilities using expert judgments, for example, by using the ordinal judgment procedure by \cite{wang13} or as outlined in~\cite{cookebook}, expect perhaps for $P_{A}(d_{2}|d_{1},a,s,e)$. The elicitation of such random probability distribution may require the defender to think about how the attacker analyzes her decision problem at $D_{2}$, 
leading to a next level of recursive thinking. There are also several ways to deal with this hierarchy of recursive analysis when eliciting such random probability distributions over {\it strategic} decisions as discussed in~\cite{rios12}.

Once these random quantities are elicited, the defender solves the insider's decision problem using backward induction. This is done following a process similar to how they solved their own decision problem  taking into account the randomness in judgments. 
First, the defender finds the random expected utility for each $d_{2} \in \mathcal{D}_{2}$
\begin{equation} \label{sim4}
\boldsymbol{\Psi_{A}}(d_{1},a,s,e) = 
\sum_{d_{2}} U_{A}(a,s,e,d_{2}) P_{A}(d_{2}|d_{1},a,s,e).
\end{equation}
Then, the defender finds the random expected utility taking into account the uncertainty 
about whether the insider is detected or not 
\begin{equation} \label{sim41}
\boldsymbol{\Psi_{A}}(d_{1},a,s) = 
\sum_{e} \Psi_{A}(d_{1},a,s,e) P_{A}(e|d_{1},a,s).
\end{equation}
\noindent Next, they find the random expected utility for each pair $(d_{1},a)$
\begin{equation} \label{sim5}
\boldsymbol{\Psi_{A}}(d_{1},a) = \sum_s 
\boldsymbol{\Psi_{A}}(d_{1},a,s) P_{A}(s|d_{1},a) ,
\end{equation}
\noindent and compute the random optimal attack $A^{*}(d_{1})$ given 
the defense $d_1$
\begin{equation}\label{sim6}
A^{*}(d_{1}) = \arg\max_{a \in \mathcal{A}} \boldsymbol{\Psi_{A}}(d_{1},a).
\end{equation}
Finally, once the defender assesses the desired conditional predictive distribution
through
\begin{equation} \label{sim7}
p_{D} (a|d_{1})  = Pr ( A^{*}(d_1) = a ) , 
\end{equation}
she is able to solve her decision problem and obtain $d_{1}^{*} \mbox{ and } d_{2}^{*}$ by solving Equations (\ref{sim1}) to (\ref{sim3}).
Similarly, $p_{D}(a|d_{1})$ can also be approximated in practice using Monte Carlo methods. In Section \ref{exe}, we illustrate this with a numerical example.


Note that in the above analysis, we have assumed that all the involved quantities are discrete.
Should some of the other quantities be continuous, 
the sums would be replaced by corresponding integrals.

\section{An ARA model incorporating Organisation Culture} \label{realistic}

\begin{figure} [!h] 
{\centering
\includegraphics [scale=0.5]{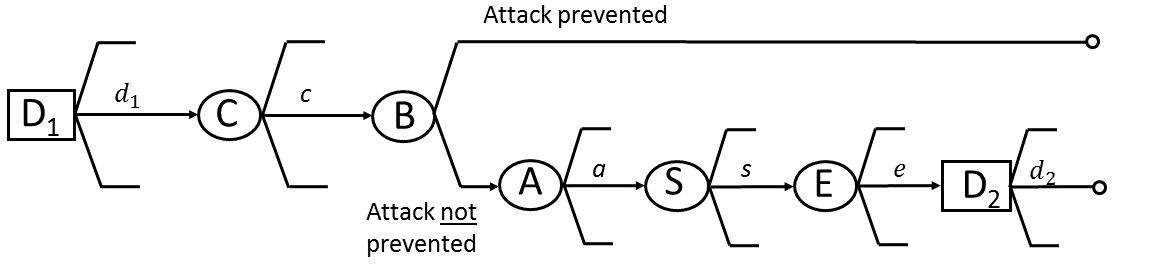}
\caption{Decision tree for the the game in Section \ref{realistic}.}

\label{id_2}
}
\end{figure}

\begin{figure} [!h] 
{\centering
\includegraphics [scale=0.45]{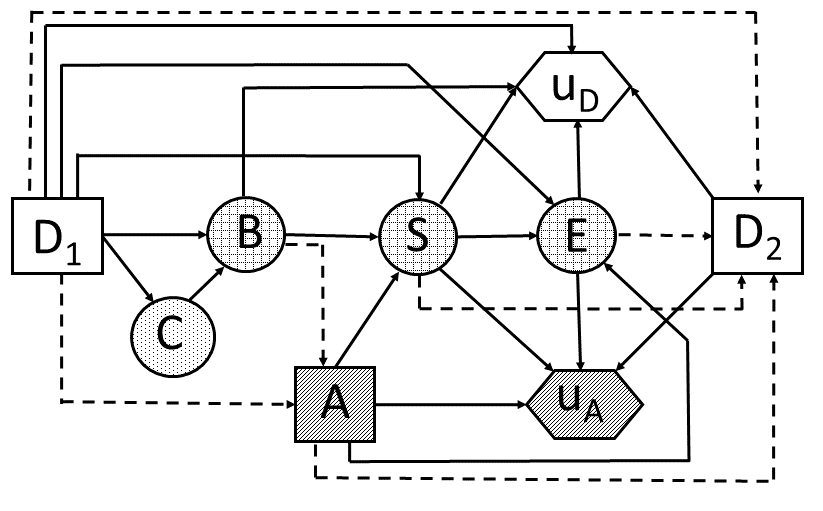}
\caption{BAID for the game in Section \ref{realistic}.} 
\label{id_21}
}
\end{figure}

The sequence of interactions between an organization and its employees
could be more complex for various reasons. Firstly, it has been described  (\cite{moore15}, \cite{liu*}, \cite{martinez08}) that the measures in $D_{1}$ can have unintended negative consequences. If the employees feel that the measures introduced 
to mitigate insider threats are intrusive or micro-managing, or even aggressive, that could lead them to react in unintended ways; this could include not
reporting suspicious activities or misusing reporting processes,
either accidentally or intentionally. At worst, it could even motivate an employee to go
rogue. Secondly, although we have treated the employees as a single entity, in reality, this group could include a large number of people and, therefore, the organization may be facing multiple actors taking different actions (mostly good, but some inadvertently harmful and some even malicious).
Usually, a majority of employees will not take any action that would harm the organization; in fact, some of them would actively help prevent insider attacks. For example, one of the possible actions 
could be to correctly follow the processes or measures set out by the organization,
possibly resulting in the successful prevention of attacks.
Finally, actions by different types of employees could be dependent (sequential) or
independent (simultaneous).
Since the number of employees and the exact nature of their actions will vary,
it is not possible to provide a general solution to the problem by modelling this 
as a multi-player game. Some of the previous research has focused on segmenting
the employees. For example,  \cite{liu*} provide a segmentation with inadvertent
and malicious insiders. Instead, we propose to model the culture
in the organisation. For simplicity, we  classify the culture in the 
organisation $C$ as \emph{good} or \emph{not so good}. We define good culture as
the one in which a majority of the employees correctly and promptly perform 
their duties including following any procedures to prevent insider attacks.
As a result there is a positive, productive and vigilant culture in the
organisation. Any suspicious activity will be promptly reported and
investigated. As a result, it will be relatively difficult to carry out an
insider attack. On the other hand, the not so good culture refers to an
environment in which, at least some of the employees will take either deliberate
or inadvertent actions creating a culture of mis-trust. 
This, in turn, could lead the employees to not feel safe to
report suspicious activities and even potentially motivate some to go rogue
and plan an insider attack. Since the culture will change with space and time,
we consider $C$ as a random event. $C$ may or may not prevent (block) and 
insider attack from going ahead. We define this as a random outcome $B,$ that
can take two values: \emph{attack prevented} or \emph{not prevented}. 
If the attack is prevented then the game ends since there is no attack. On the other hand, if it is not prevented, then the attacker $A$ will carry out an attack that will result in an outcome from the set $\mathcal{S}.$ The game will conclude by observing events $E$ and $D_{2}$ as defined in Section \ref{kaka}. The decision tree for this game is shown in Figure \ref{id_2} and the BAID (Bi-agent influence diagram) in Figure \ref{id_21}. 

To find the ARA solution for this game, the defender must first quantify the following.
\begin{enumerate}
\item{Her predictive distribution $p_{D}(c|d_{1})$ about the 
culture in the organisation given the implemented defenses $d_{1}$.}
\item{Her predictive distribution $p_{D}(b|d_{1},c)$ about the outcome of
such a culture (whether it can prevent an insider attack or not), given $c$ and $d_{1}$.}
\item{Her predictive distribution $p_{D}(a|d_{1}, b=\mbox{not prevented})$ about the attack
that will be chosen at node $A$ given the outcome $b$ was \emph{not prevented} and action $d_{1}$.}
\item{Her predictive distribution $p_{D}(s|a,d_{1},b=\mbox{not prevented})$
about the outcome  of the attack, given  the attack $a$, the defenses $d_1$ implemented and given that the outcome $b$ was \emph{not prevented}.}
\item{Her predictive distribution $p_{D}(e|d_{1},a,s)$ about the detection 
of the attacker, given the actions $d_{1}$ and $a$ and the outcome $s.$}
\item{The utility function $u_{D}(d_{1},b,s,e,d_{2})$ given their first 
and second defensive actions, the outcome $b$ of the organisational culture,
the outcome $s$ of the attack and the eventual detection $e$ of the attacker.}
\end{enumerate}
Given these, the defender works backwards along the decision tree in Figure \ref{id_2}.
First, she seeks to find her optimal second defensive action $d_{2}^{*}(d_{1},b,s,e)$ by maximizing her utility
\begin{equation} \label{sim72}
d_{2}^{*}(d_{1},b,s,e)
= \arg\max_{d_{2}\in \mathcal{D}_{2}} u_{D}(d_{1},b,s,e,d_{2}).
\end{equation}
Then, for each $(d_{1},b,a ) \in \mathcal{D}_{1}\times\mathcal{B} \times \mathcal{A},$
she seeks to compute her expected utility 
by taking into account the uncertainty in $E$ and $S$, as follows
\begin{equation} \label{sim8}
\psi_{D}(d_{1},b,a) = \sum_s \left( \sum_{e} u_{D}(d_{1},b,s,e,d_{2}^{*} )
p_{D}(e|d_{1},a,s) \right )\, p_{D}(s|d_1,b,a ) ; 
\end{equation}
next she computes her expected utility $\psi_{D}(d_{1},b)$  by considering the uncertainty in $A$, 
through 
\begin{equation} \label{sim9}
\psi_{D}(d_{1},b) =  \sum_a \psi_{D}(d_{1},b,a) p_{D}(a|d_{1},b) , 
\end{equation}
and, then, integrate out the uncertainty in $b$ and $c$
through 
\begin{equation}  \label{sim10}
\psi_{D}(d_{1}) =  \sum_c \left ( \sum_b \psi_{D}(d_{1},b) p_{D}(b|d_{1},c) \right) p_{D}(c|d_{1}) ,
\end{equation}
to find her expected utility $\psi_{D}(d_{1})$ for each $d_{1}.$ Finally, she can 
find her initial decision of maximum utility 
through $d_{1}^{*} = \arg\max_{d_{1}\in \mathcal{D}_{1}}  \psi_{D}(d_{1}).$
This backward induction shows that the defender's optimal strategy is to first choose $d_{1}^{*}$ and then, after having observed $b,a, s$ and $e$, choose action $d_{2}^{*}(d_{1}^{*},b,s,e).$

The above analysis requires the defender to elicit $p_{D}(a|d_{1},b)$,  
which is not straightforward even when the attack could only go ahead if $b=\mbox{not prevented}$. This is due to its strategic nature. The defender could model the insider's strategic analysis process by assuming that he will perform an analysis similar to hers to find
his optimal attack $A^{*}$. 
To elicit it, the defender must assess
$U_{A}(a,s,e,d_{2})$,
$P_{A}(s|d_1,b,a)$,
$P_{A}(e| d_1,a,s)$ and 
$P_{A}(d_{2}|d_{1},a,s,e)$, as discussed in Section \ref{kaka}.
Once elicited, the defender solves the attacker's decision problem using
backward induction, similar to how they solved their own decision problem.
First, the defender integrates out
action $d_{2}$ 
computing the random expected utilities
\begin{equation} \label{sim12}
{\bf \Psi_{A}}(d_{1},a,s,e) = \sum_{d_{2}} U_{A}(a,s,e,d_{2}) P_{A}(d_{2}|d_{1},a,s,e).
\end{equation}
\noindent Then, she finds her random expected utilities for the attacker
taking into account the uncertainty in $e,$
\begin{equation} \label{sim12}
{\bf \Psi_{A}}(d_{1},a,s) = \sum_{e} {\bf \Psi_{A}}(d_1, a,s,e) P_{A}(e|d_{1},a,s).
\end{equation}
\noindent Finally, they integrate out $s  \in \mathcal{S}$ to obtain
\begin{equation} \label{sim13}
{\bf \Psi_{A}}(d_{1},a, b ) = \sum_s {\bf \Psi_{A}}(a,d_1,s) P_{A}(s|a, d_1, b)  ,
\end{equation}
\noindent and compute the attacker's random optimal action 
\begin{equation} \label{sim14}
A^{*} (d_1, b) = \arg\max_{a \in \mathcal{A}} {\bf \Psi_{A}}(d_{1},a, b).
\end{equation}
This produces the defender's desired predictive distribution about the attack  
\begin{equation} \label{sim15}
p_{D}(a|d_{1},b)= Pr ( A^*(d_1, b) = a ) 
\end{equation}


Again, as in Section \ref{kaka}, we have assumed that all quantities are discrete. Should some of the quantities be continuous, the corresponding sums would be replaced by integrals. In practice, $d_{1}^{*}, d_{2}^{*} \mbox{ and } P_{D}(a|d_{1},b )$ can be obtained using Monte Carlo methods, as shown in Section \ref{ARA}. In Section \ref{exe}, we illustrate this using a numerical example.

\section{Example} \label{exe}

We consider an insider threat scenario motivated by \cite{martinez08} in which the malicious insider attempts to harm the incumbent organization without getting caught.
The organization focuses on information/data collection and 
needs to protect itself. 
It already has its sites and IT systems protected so that only authorized personnel are able to access them.  However, anticipating attacks, the organization is considering implementing an additional security layer to defend itself. The defensive actions in $D_{1}$ under consideration are
\begin{enumerate}
\item anomaly detection/data provenance tools;  
\item information security measures and employee training; and 
\item carrying out random audits.
\end{enumerate}
The malicious insider's aim could be financial fraud, data theft, espionage or whistle blowing. Regardless of the exact nature of the attack, we assume that the attacker's 
options in $A$ refer to its scale, say \emph{small}, \emph{medium} or \emph{large}. For simplicity, we assume that the attack will either fully succeed or
fail at $S$. Once the attack has been carried out, the attacker may or may not be detected, an event represented by $E$. Upon observing $E,$ the organization can choose to carry out one of the following 
defensive actions in $D_{2}$:
\begin{enumerate}
\item major upgrade of defenses;
\item minor upgrade of defenses; or
\item no upgrade.
\end{enumerate}
\subsection{Solution using the defend-attack-defend model} \label{ex_simple}
We first analyze the problem using the model in Section \ref{kaka}. 
We start by assessing the defender's utility function $u_{D}(d_{1},s,e,d_{2}).$
We assume for simplicity that such function additively aggregates the marginal utilities associated, respectively, with her initial defense action $d_1$, the outcome of the attack and whether this is detected or not $(s,e)$, and her recovery action $d_{2}$ through 
\begin{equation} \label{ud_def}
u_{D}(d_{1},s,e,d_{2}) = u(d_{1}) + u(s,e) + u(d_{2}) .
\end{equation}
These marginal utilities are given in Tables \ref{Ud} and \ref{Uas}. They are estimated in a common scale from -100 to 100.
The utility $u_{D}$ can then be computed for each
combination; for example, 
$u_{D}(d_{1}= \mbox { {\em random audit}}, s = \mbox{ {\em not successful}}, e = \mbox{{\em detected}}, d_{2}= \mbox{\em no upgrade}) = -50 + 100 - 100 = -50.$
Observe that while $u(d_{1})$ measures the relative utilities of the monetary costs associated with each initial defensive action (therefore the negative values), $u(d_{2})$ will, in itself, combine the defender's cost and expected benefits associated with her recovery action $d_2$. This amounts to considering her expected utility for the consequences after her decision at $D_{2}$. For instance, in this example, the defender seems to think that Minor upgrade will be the best recovery action $d_{2}$ considering the investment cost and the expected benefits. In real life, $u(d_{2})$ will likely depend on $s$ and $e$, that is, it would be $u(d_{2}|s,e).$ Because, $u(\mbox{\emph{Major upgrade}}| \mbox{\emph{ Successful, Not detected}})$ will likely be different from $u(\mbox{\emph{Major upgrade}}| \mbox{\emph{ Not Successful, Detected}}),$ for example. However, here, for the sake of simplicity we do not consider this dependence when eliciting $u(d_{2}).$
\begin{table}[hbt]
\renewcommand{\arraystretch}{1.2}
\caption{\label{Ud} Marginal utilities associated with defensive actions $d_{1}$ and $d_{2}$.}
\centering
\begin{tabular}{|l|r|l|r|}\hline
$d_{1}$ & $u(d_{1})$ & $d_{2}$ & $u(d_{2})$ \\\hline
Anom. det. \& Data prov. & -100  & Major upgrade & 0\\
Info. Sec.\& train.  & -60 & Minor upgrade & 25\\
Random audits & -50 & No upgrade & -100\\ \hline
 \end{tabular}
\end{table}
\begin{table}[hbt]
\renewcommand{\arraystretch}{1.2}
\caption{\label{Uas} Marginal utility for every $(s,e)$ combination.}
\centering
\begin{tabular}{|l|l|r|}\hline
$s$ & $e$ & $u(s,e)$ \\\hline
 Success & Yes & 50\\
 Success & No & -100\\
 Fail & Yes & 100\\
 Fail & No & 0 \\ \hline
 \end{tabular}
\end{table}
 In order to implement the ARA solution, the defender must first identify the action $d_{2}^{*}(d_{1},s,e)$  maximizing their utility. 
 In this case, $d_{2}^{*}$ turns out to be `Minor upgrade'.
 
 Then, she must compute her expected utility $\psi_{D}(d_{1},a)$ using Eq.~(\ref{sim2}). This requires $p_{D}(e|d_{1},a,s)$ and $p_{D}(s|d_{1},a)$. Suppose that her elicited probabilities $p_{D}(e|d_{1},a,s)$ are as shown in Table~\ref{P_e_d1_a_s}. We can see that the defender believes that there is a higher probability of detecting the attacker if $d_1=$ {\em Anomaly detection and data provenance systems} and the larger the attack is. 
 Suppose also that elicited the probabilities $p_{D}(s|d_{1},a)$ are as shown in Table \ref{P_s_d1_a}, with probabilities of failed attacks obtained  through $p_{D}(\mbox{not successful }|d_{1}, a) =$ $1- p_{D}(\mbox{successful }|d_{1}, a).$
 The expected utilities $\psi_{D}(d_{1},a)$  can now be computed, producing the values in Table~\ref{Psi_d1_a}.
 
Now, to compute the defender's expected utility associated with each $d_{1} \in D_{1}$ using Eq.~(\ref{sim3}), we need the predictive distribution $p_{D}(a|d_{1})$ about what the malicious insider may do. 
  Assume first that the defender has elicited $p_{D}(a|d_{1})$ using her own beliefs as in Table \ref{P_a_d1}.
  The defender's expected utility $\psi_{D}(d_{1})$ for each action $d_{1}$  is computed;  for example, $\psi_{D}(\mbox{\em random audits}) = -62.5 \times 0.5 - 53 \times 0.4 -43.5 \times 0.1 = -56.8.$ 
Similarly, the expected utility for {\em anomaly detection and data provenance}
is $-9.9$, whereas for {\em information security and training} is $-7.6.$ This implies that the optimal option for the 
organization is to invest in {\em information security and staff training}; 
moreover, if the attack was to happen, then, irrespective
of whether it was successful or not, the optimal follow-up action would be {\em to carry on minor upgrades}.  
 
 \begin{table}[!t]
\renewcommand{\arraystretch}{1.2}
\caption{\label{P_e_d1_a_s} Defender's elicited probabilities $p_{D}(E = \mbox{Yes}|d_{1},a,s)$  for every $(d_{1},a,s)$.} 
\centering
\begin{tabular}{|l|c|c|c|c|c|c|c|}\hline 
 & \multicolumn{3}{c}{$S=$ Success} & \multicolumn{3}{|c|}{$S=$ Fail}\\ \hline
 & \multicolumn{3}{c|}{$A=$}  &  \multicolumn{3}{c|}{$A=$} \\
$D_1$  & Small & Med. & Large & Small & Med. & Large\\ \hline
Anom.det. \&  prov. & $0.6$ & $0.7$ & $0.8$ & $0.7$ & $0.8$ & $0.9$ \\
Info.sec. \& train. & $0.3$ & $0.4$ & $0.5$ & $0.4$ & $0.5$ & $0.6$\\
Random audits & $0.1$ & $0.1$ & $0.1$ & $0.1$ & $0.1$ & $0.1$\\ \hline
 \end{tabular}
\end{table}\textbf{}

\begin{table}[!t]
\renewcommand{\arraystretch}{1.2}
\caption{\label{P_s_d1_a} Defender's elicited probabilities $p_{D}(S= \mbox{successful }|d_{1}, a)$ for every $(d_1 , a)$.} 
\centering
\begin{tabular}{|l|r|r|r|}\hline
$D_{1}$ & $A=$ small & $A=$ med. & $A=$ large\\\hline
Anom. det. \& Data prov. & 0.1 & 0.07 & 0.05\\
Info. sec. \& train. & 0.3 & 0.25 & 0.2\\
Random audits & 0.5 & 0.4 & 0.3\\ \hline
 \end{tabular}
\end{table}

\begin{table}[!t]
\renewcommand{\arraystretch}{1.2}
\caption{\label{Psi_d1_a} Computed $\psi_{D}(d_{1}, a)$ for every $(d_{1}, a)$ combination.} 
\centering
\begin{tabular}{|l|r|r|r|}\hline
$D_{1}$ & $A=$ small & $A=$ med. & $A=$ large\\\hline
Anom. det. \& Data prov. & -13 & -0.25 & 11.5\\
Info. sec. \& train.& -23.5 & -7.5 & 8\\
Random audits & -62.5 & -53 & -43.5\\ \hline
 \end{tabular}
\end{table}

\begin{table}[!t]
\renewcommand{\arraystretch}{1.2}
\caption{\label{P_a_d1} $p_{D}(a|d_{1})$ directly elicited by defender without using ARA.} 
\centering
\begin{tabular}{|l|r|r|r|}\hline
$D_{1}$ & $A=$ small & $A=$ med. & $A=$ large\\\hline
Anom. det. \& Data prov. & 0.8 & 0.15 & 0.05\\
Info. sec. \& train. & 0.2 & 0.6 & 0.2\\
Random audits & 0.5 & 0.4 & 0.1\\
\hline
 \end{tabular}
\end{table}

\begin{table}[!t]
\renewcommand{\arraystretch}{1.2}
\caption{\label{Ua_a_s_e_d2} Attacker's random utilities $U_{A}(a,s,e,d_{2})$ elicited by the defender.} 
\centering
\begin{footnotesize}
\begin{tabular}{|l|c|c|c|c|}\hline 
 $A=$ small & \multicolumn{2}{c}{$E=$ Yes} & \multicolumn{2}{c}{$E=$ No}\\ \hline
 $D_2$ & Succ. & Fail & Succ. & Fail \\ \hline
Maj.upgr. & $N(-85,3)$ & $N(-95,1)$ & $N(-80,5)$ & $N(-90,2)$ \\
Min.upgr. & $N(-55,7)$ & $N(-65,3)$ & $N(-50,10)$ & $N(-60,5)$ \\
No upgr. & $100 - Exp(3)$ & $100 - Exp(3)$ & $100 - Exp(5)$ & $100 - Exp(5)$ \\
\hline
 $A=$ medium & \multicolumn{2}{c}{$E=$ Yes} & \multicolumn{2}{c}{$E=$ No}\\ \hline
Maj.upgr. & $N(-85,3)$ & $N(-95,1)$  & $N(-80,5)$ & $N(-90,2)$\\
Min.upgr. & $N(-50,5)$ & $N(-65,3)$ & $N(-40,10)$ & $N(-60,5)$\\
No upgr. & $100 - Exp(2)$ & $100 - Exp(2)$  & $100 - Exp(3)$ & $100 - Exp(3)$\\
\hline
 $A=$ large & \multicolumn{2}{c}{$E=$ Yes} & \multicolumn{2}{c}{$E=$ No}\\ \hline
Maj.upgr. & $N(-85,3)$ & $N(-95,1)$  & $N(-80,5)$ & $N(-90,2)$\\
Min.upgr. & $N(-20,5)$ & $N(-65,3)$  & $N(-30,10)$ & $N(-60,5)$\\
No upgr. & $100 - Exp(1)$ & $100 - Exp(1)$  & $100 - Exp(1)$ & $100 - Exp(1)$\\
\hline
 \end{tabular}
 \end{footnotesize}
\end{table}

\begin{table}[!t]
\renewcommand{\arraystretch}{1.2}
\caption{\label{Pa_d2_d1_a_s_e} Attacker's random probabilities $P_{A}(d_{2}|d_{1},a,s,e)$ elicited by the defender. Dirichlet distributions with first component corresponding to {\em Major upgrade}, second to {\em Minor upgrade} and third to {\em No upgrade}.} 
\centering
\begin{footnotesize}
\begin{tabular}{|l|c|c|c|c|}\hline
 $A=$ small & \multicolumn{2}{c}{$E=$ Yes} & \multicolumn{2}{c|}{$E=$ No}\\ \hline
 $D_1$ & Succ. & Fail & Succ. & Fail \\ \hline
Anom. det. \& Data prov. & $Dir(1,3,6)$ & $Dir(1,9,90)$ & $Dir(2,2,4)$ & $Dir(2,18,80)$\\
Info. sec. \& train. & $Dir(1,5,4)$  & $Dir(1,9,90)$ & $Dir(2,6,2)$ & $Dir(2,18,80)$\\
Random audits & $Dir(1,4,5)$  & $Dir(1,9,90)$ & $Dir(2,5,3)$ & $Dir(2,18,80)$\\ 
\hline
 $A=$ Medium & \multicolumn{2}{c}{$E=$ Yes} & \multicolumn{2}{c|}{$E=$ No}\\ \hline
$D_1$ & Succ. & Fail & Succ. & Fail \\ \hline
Anom. det. \& Data prov. &$Dir(2.5,7,0.5)$  & $Dir(1,9,90)$ & $Dir(3,6.5,0.5)$ & $Dir(2,18,80)$\\
Info. sec. \& train. & $ Dir(1.5,8,0.5)$  & $Dir(1,9,90)$ & $Dir(2,7.5,0.5)$ & $Dir(2,18,80)$\\
Random audits & $Dir(2,6,2)$  & $Dir(1,9,90)$ & $Dir(3,7,1)$ & $Dir(2,18,80)$\\
\hline
 $A=$ Large & \multicolumn{2}{c}{$E=$ Yes} & \multicolumn{2}{c|}{$E=$ No}\\ \hline
$D_1$ & Succ. & Fail & Succ. & Fail \\ \hline
 Anom. det. \& Data prov. & $Dir(5,4.9,0.1)$  & $Dir(1,9,90)$ & $Dir(5.5,4.4,0.1)$ & $Dir(2,18,80)$\\
 Info. sec. \& train. & $Dir(5,4.9,0.1)$  & $Dir(1,9,90)$ & $Dir(5.5,4.4,0.1)$ & $Dir(2,18,80)$\\
 Random audits & $Dir(5,4.9,0.1)$  & $Dir(1,9,90)$&  $Dir(5.5,4.4,0.1)$ & $Dir(2,18,80)$\\
\hline
 \end{tabular}
 \end{footnotesize}
\end{table}

\begin{table}[!t]
\renewcommand{\arraystretch}{1.2}
\caption{\label{Pa_e_d1_a_s} Attacker’s random probabilities $P_{A}(E= \mbox{Yes}|d_{1},a,s)$ elicited by the defender. Note that $P(E= \mbox{No}|d_{1},a,s) = 1 - P(E= \mbox{Yes}|d_{1},a,s)$} 
\centering
\begin{footnotesize}
\begin{tabular}{|l|c|c|}\hline
 $A=$ small \\ \hline
 $D_1$ & Succ. & Fail  \\ \hline
Anom. det. \& Data prov. & $Beta(6,4)$ & $Beta(7,3)$ \\
Info. sec. \& train. & $Beta(3,7)$  & $Beta(4,6)$ \\
Random audits & $Beta(1,9)$  & $Beta(1,9)$ \\ 
\hline
 $A=$ Medium \\ \hline
$D_1$ & Succ. & Fail \\ \hline
Anom. det. \& Data prov. &$Beta(7,3)$  & $Beta(8,2)$ \\
Info. sec. \& train. & $ Beta(4,6)$  & $Beta(5,5)$ \\
Random audits & $Beta(1,9)$  & $Beta(1,9)$ \\
\hline
 $A=$ Large \\ \hline
$D_1$ & Succ. & Fail \\ \hline
 Anom. det. \& Data prov. & $Beta(8,2)$  & $Beta(9,1)$ \\
 Info. sec. \& train. & $Beta(5,5)$  & $Beta(6,4)$ \\
 Random audits & $Beta(1,9)$  & $Beta(1,9)$\\
\hline
 \end{tabular}
 \end{footnotesize}
\end{table}

\begin{table}[!t]
\renewcommand{\arraystretch}{1.2}
\caption{\label{PA_s_d1_a} Attacker’s random probabilities $P_{A}(S= \mbox{successful }|d_{1}, a)$ elicited by defender for every $(d_1, s)$ combination.} 
\centering
\begin{tabular}{|l|r|r|r|}\hline 
$D_{1}$ & $A=$ small & $A=$ medium & $A=$ large\\\hline
Anom.\ det. \& Data prov. & $Beta(4,6)$ & $Beta(2,8)$ & $Beta(0.5, 9.5)$\\
Inf.sec. \& train. & $Beta(9,1)$ & $Beta(8,2)$ & $Beta(7,3)$\\
Random audits & $Beta(7,3)$ & $Beta(6,4)$ & $Beta(3,7)$\\ \hline
 \end{tabular}
\end{table}

\begin{table}[!t]
\renewcommand{\arraystretch}{1.2}
\caption{\label{PA_a_d1} Predictive attack probabilities $p_{D}(a|d_{1})$ computed by the defender using ARA} 
\centering
\begin{tabular}{|l|r|r|r|}\hline 
$D_{1}$ & $A=$ small & $A=$ med. & $A=$ large\\\hline
Anom.det. \& Data prov. & 0.067 & 0.090 & 0.843\\
Info.sec. \& train. & 0.499 & 0.189 & 0.312\\
Random audits & 0.196 & 0.105 & 0.699\\
\hline
 \end{tabular}
\end{table}
 
We now illustrate how $p_{D}(a|d_{1})$ could be obtained by modeling the attacker's strategic analysis process using Eqs.~(\ref{sim4}) to (\ref{sim7}). 
To do this, the defender must elicit the attacker's random utilities and probabilities $U_{A}(a,s,e,d_{2})$, $P_{A}(e|d_{1},a,s)$, $P_{A}(s|a,d_{1})$ and $P_{A}(d_{2}|d_{1},a,s,e)$,  
using any information the defender might have,  as well as considering the possible motivations for the attackers and their skill level. 
To elicit the distribution of the random utilities, $U_{A}(a,s,e,d_{2}),$ the defender will try to think from the point of view of the attacker. She may think that the attacker is likely to strongly prefer $d_{2}$ = \emph{no upgrade} since this means
 that future attacks will remain equally likely to succeed as now whereas she thinks that $d_{2}$ = \emph{major upgrade} would be perceived by the attacker as strongly undesirable since
 that will make any future attacks less likely to succeed. Also, the attacker is better off not being detected and succeeding in his  
 attack. Table \ref{Ua_a_s_e_d2} shows the attacker's random utilities $U_{A}(a,s,e,d_{2})$ elicited by the defender, scaled between $-100$ and $+100$.

 When eliciting $P_{A}(d_{2}|d_{1},a,s,e)$, the defender may think that the attacker believes that her decision is 
 influenced by the costs involved in establishing  the upgrades in $D_{2}$ as well as the magnitude and success of the attack.
 Thus, for example, from her perspective, he may believe that she is more likely to invest in a {\em Major upgrade} in the wake of a large, successful attack, but more likely
 to go with {\em minor upgrades} or {\em no upgrade} in all other circumstances. Table \ref{Pa_d2_d1_a_s_e} shows the distributions
 elicited for $P_{A}(d_{2}|d_{1},a,s,e)$ by the defender. For each of the Dirichlet distributions, the first component of the
 triplet corresponds to {\em Major upgrade}, the second to {\em Minor upgrade} and, finally, the third one to {\em No upgrade}.  For example, since
 the defender thinks that an upgrade is considered (by the attacker) to be less likely in the event of a failed attack, she could elicit
 a $Dir(1,9,90)$ distribution for it. On the other hand, the $Dir(5,4.9,0.1)$ indicates that an
 upgrade is considered almost certainly likely in the wake of a successful large attack. 
 
  The defender then elicits the distributions for $P_{A}(e|d_{1},a,s)$, as shown in  Table~\ref{Pa_e_d1_a_s}. She thinks that the attacker believes that the probability of being detected is highest if  $D_{1}$ = {\em anomaly detection \& data provenance} and lowest if $D_{1}$ = {\em random audits} and also is higher in the wake of a failed attack versus a successful one and higher still the larger the attack was.
 Finally, Table \ref{PA_s_d1_a} shows the distributions elicited for $P_{A}(s|d_{1}, a)$ by the defender. For example, she believes that the attacker thinks that an attack is much more likely to succeed  if $D_{1}$ is {\em information security and training} compared to the other options. Also, she believes that the attacker thinks that a small attack is much more likely to succeed than a medium or large attack.

  Given these assessments, the defender can compute her estimate of $p_{D}(a|d_{1})$ by Monte Carlo simulation.
 For each combination of $(d_{1}, a, s, e)$, we simulate $N= 1000$ samples from the attacker's random utilities $U_{A}(a,s,e,d_{2})$ and probabilities $P_{A}(d_{2}|d_{1},a,s,e)$ to obtain samples of the attacker's  expected utility ${\bf \Psi_{A}}(d_{1},a,s,e)$ using Eq. (\ref{sim4}). Then, we simulate samples from $P_{A}(e|d_{1},a,s)$ to obtain samples of the attacker's expected utility ${\bf \Psi_{A}}(d_{1},a,s)$ using Eq. (\ref{sim41}). Finally, we simulate samples from $P_{A}(s|d_{1}, a)$ to obtain samples of ${\bf \Psi_{A}}(d_{1},a)$ using Eq. (\ref{sim5}). Then, for each of the samples, we find the optimal attack $a$ maximizing ${\bf \Psi_{A}}(d_{1},a)$ and, estimate $p_{D}(a|d_{1})$ by counting how  many times (out of $N$) the optimal attack corresponds to each particular $a \in \cal{A}$ for each given $d_{1}.$ The estimated probabilities are presented in Table \ref{PA_a_d1}. 
 
 We can now use these estimates of $p_{D}(a|d_{1})$ to compute the defender's expected utility $\psi_{D}(d_{1})$ using 
 Eq.~(\ref{sim3}). The expected utility for the {\em anomaly detection and data provenance} comes out to be $8.8$;  
 for {\em information security and training}, $-10.65,$;  and, finally,
 for the  {\em random audits},  $-48.22.$
 This implies that,  in this case, the optimal initial defense for the organization is to invest in {\em anomaly detection and data provenance} and,
 if the attack was to happen irrespective  of whether it was successful or not, the optimal follow-up action would be {\em to carry on minor upgrades} of their existing defenses. 
 Of course, this is based on the elicited inputs given by the defender.
Moreover, observe that the $p_{D}(a|d_{1})$ assessed by defender by modeling the  attacker's strategic analysis (Table \ref{PA_a_d1}), that is using ARA, turns out to be quite
 different from the one elicited directly by the defender without using 
 ARA but based only on her first uninstructed intuition (Table \ref{P_a_d1}), which in turn leads to a different optimal defense $d_{1}.$

\subsection{Solution using the organisational culture model} \label{ex_real}

We now analyze this problem using the model discussed in Section \ref{realistic} by incorporating the organisational culture. Again, we start by assessing the defender's utility function $u_{D}(d_{1},b,s,e, d_{2}).$ Just like we did in Section \ref{ex_simple}, we assume that  $u_{D}$ adopts the form 
\begin{equation}
u_{D}(d_{1},b,s,e,d_{2}) = u(d_{1}) + u(b) + u(s,e) + u(d_{2}),
\end{equation}
where $u(d_{1})$ and $u(d_{2})$ are as defined in Table \ref{Ud} and $u(s,e)$ is the
same as in Table \ref{Uas}. 
Suppose the organisation elicits $u(b)$ by assigning $u(b=\mbox{blocked}) = 100$ and $u(b=\mbox{not blocked}) = -50$, representing, in a $-100$ to $+100$ scale, the relative utilities for the organisation associated with a blocked insider attack (very highly desirable outcome) and with an attack that could not be blocked (a considerably 
undesirable outcome), respectively. 
We calculate $u_{D}(d_{1},b,s,e,d_{2})$ using Tables \ref{Ud} and \ref{Uas}.

In order to implement the ARA solution to this problem, the defender must use backward induction. She first finds her action $d_{2}^{*}(d_{1},b,s,e)$ maximizing her utility. In this case, again, $d_{2}^{*}$ turns out to be `Minor upgrade', as it maximizes the term $u(d_{2})$ in $u_{D}$. Next, we need to elicit the probabilities $p_{D}(e|d_{1},a,s).$ Suppose that these are same elicited in Table \ref{P_e_d1_a_s}. Next, to integrate out the uncertainty in $s$, we need to elicit the probability $p_{D}(s|d_1,b,a ).$ Note, however, as illustrated in Figure \ref{id_2}, that the attack will only proceed if it was not blocked from taking place beforehand. Therefore, at this stage we only need to elicit $p_{D}(s|d_1,b = \mbox{not blocked},a)$, given that the attack was not blocked, which we suppose is the same as $p_{D}(s|d_1,a)$ in Table~\ref{P_s_d1_a}. We can now compute $\psi_{D}(d_{1},b,a)$ using Equation (\ref{sim8}), see Table \ref{psiD_d1_b_a}.

\begin{table}[!t]
\renewcommand{\arraystretch}{1.2}
\caption{\label{psiD_d1_b_a} $\psi_{D}(d_{1},b,a)$ for when $b = \mbox{not blocked}$.} 
\centering
\begin{tabular}{|l|c|c|c|}\hline 
 & \multicolumn{3}{c|}{$A$} \\ \hline
 $D_1$ & Small & Med. & Large \\ \hline
Anom.det. \&  prov. & $-88$ & $-75.25$ & $-63.5$ \\
Info.sec. \& train. & $-98.5$ & $-82.5$ & $-67$ \\
Random audits & $-137.5$ & $-128$ & $-118.5$ \\
\hline
 \end{tabular}
\end{table}

 We next seek to compute the expected utility $\psi_{D}(d_{1},b)$ using Equation (\ref{sim9}), which requires us to elicit $p_{D}(a|d_{1},b).$ This can be elicited either directly using the defender's domain knowledge or indirectly by modeling the malicious insider's strategic analysis process using Equations (\ref{sim12}) to (\ref{sim15}). Again, note that the attack could only go ahead if it was not blocked beforehand. Therefore, what we really need to elicit is $p_{D}(a|d_{1},b= \mbox{not blocked}).$ Thus, suppose that given that the attack was not blocked, $p_{D}(a|d_1,b = \mbox{not blocked})$ is the same as $p_{D}(a|d_1)$ in Table \ref{P_a_d1} when eliciting these probabilities directly and the one in Table \ref{PA_a_d1} when eliciting them using ARA. This gives us $\psi_{D}(d_{1},b = \mbox{not blocked}).$ For the case where the attack was successfully blocked in advance, we have $\psi_{D}(d_{1},b = \mbox{blocked}) = u(b = \mbox{blocked})+u(d_{1}).$ Tables \ref{psiD_d1_b} and \ref{psiD_d1_b1} show the computed values of $\psi_{D}(d_{1},b)$ using both $p_{D}(a|d_{1},b)$ expressions obtained using direct
 elicitation and ARA respectively. 

\begin{table}[!h]
\renewcommand{\arraystretch}{1.2}
\caption{\label{psiD_d1_b} Computed values for $\psi_{D}(d_{1},b)$ using $p_{D}(a|d_{1},b)$ in Table \ref{P_a_d1}.} 
\centering
\begin{tabular}{|l|c|c|}\hline 
 $D_1$ & $b = \mbox{blocked}$ & $b = \mbox{not blocked}$\\ \hline
Anom.det. \&  prov. & $0$ & $-59.86$ \\
Info.sec. \& train. & $40$ & $-57.6$ \\
Random audits & $50$ & $-106.8$ \\
\hline
 \end{tabular}
\end{table}

\begin{table}[!h]
\renewcommand{\arraystretch}{1.2}
\caption{\label{psiD_d1_b1} Computed values for $\psi_{D}(d_{1},b)$ using $p_{D}(a|d_{1},b)$ in Table \ref{PA_a_d1}.} 
\centering
\begin{tabular}{|l|c|c|}\hline 
 $D_1$ & $b = \mbox{blocked}$ & $b = \mbox{not blocked}$\\ \hline
Anom.det. \&  prov. & $0$ & $-36.04$ \\
Info.sec. \& train. & $40$ & $-22.54$ \\
Random audits & $50$ & $-192.46$ \\
\hline
 \end{tabular}
\end{table}

Finally, the defender needs to integrate out the uncertainties regarding whether the attack will be blocked in advance and regarding the culture in the organisation that the insider is likely to come across using Equation (\ref{sim10}). This requires eliciting $p_{D}(b|d_{1},c)$ and $p_{D}(c|d_{1}).$ Suppose the defender believes that the chances of the attack being blocked in advance are very high in presence of a good culture in the organisation. Also that both the information security, as well as the random audit options, are only really effective at preventing an attack if the culture in the organisation is good. Table \ref{pD_b_d1_c} shows these probabilities. Eliciting $p_{D}(c|d_{1})$ can be relatively straightforward and, unlike most of the probabilities and utilities displayed so far for this analysis, these can be elicited by using real data collected through employee surveys, for example. Suppose that the organisation concludes that a majority of its employees would prefer to be trusted and treated as partners. Also that the employees are likely to be unhappy if they feel as if they are under surveillance or subject to random checks. Suppose these elicited probabilities $p_{D}(c|d_{1})$ are as shown in Table \ref{pD_b_d1_c}. 

\begin{table}[!h]
\renewcommand{\arraystretch}{1.2}
\caption{\label{pD_b_d1_c} Elicited $p_{D}(b = \mbox{blocked}|d_{1},c)$ and $p_{D}(c=\mbox{good}|d_{1})$.} 
\centering
\begin{tabular}{|l|c|c|c|}\hline 
& \multicolumn{2}{c|}{$p_{D}(b = \mbox{blocked}|d_{1},c)$} & $p_{D}(c = \mbox{good}|d_{1})$ \\ \hline
 $D_1$ & $c = \mbox{good}$ & $c = \mbox{not so good}$ & \\ \hline
Anom.det. \&  prov. & $0.9$ & $0.7$ & $0.5$ \\
Info.sec. \& train. & $0.8$ & $0.1$ & $0.7$\\
Random audits & $0.7$ & $0.2$ & $0.3$\\
\hline
 \end{tabular}
\end{table}

 Then, using Equation \ref{sim10} and the probabilties $p_{D}(a|d_{1},b)$ in Table \ref{P_a_d1}, we get that  $\psi_{D}(\mbox{Anom.det. \&  prov.}) = -11.97, \, \psi_{D}(\mbox{Info.sec. \& train.}) = -0.02$ and $\psi_{D}(\mbox{Random audits}) = -51.92.$ Based on these elicited utilities and probabilities, the optimal defensive action $d_1^*$ is to {\em invest in Information security} and {\em training of the staff}.

Similarly, when we use the ARA probabilities $p_{D}(a|d_{1},b)$ in Table \ref{PA_a_d1}, we have that   
$\psi_{D}(\mbox{Anom.det. \&  prov.}) = -7.21, \, \psi_{D}(\mbox{nfo.sec. \& train.}) = 14.36$ and $\psi_{D}(\mbox{Random audits}) = -107.6.$ Therefore, based on the elicited utilities and probabilities, the optimal defensive action $d_1^*$ is {\em invest in Information security} and {\em training of the staff}. Note, however, that in this case the expected utilities $\psi_{D}$ are quite different than before.

\subsection{Model uncertainty}

We have seen how one can find the ARA optimal solution for the organization given a specific game/model. The expected utility $\psi_{D}(d_{1})$ that we find with each of the models is, in fact, $\psi_{D}(d_{1}|M),$ where  $M$ refers to the game under consideration. In reality though, the exact scenario will be unknown. 
A Bayesian approach allows the organization to incorporate model uncertainty into the analysis and compute the expected utility taking into account 
model uncertainty concepts as in the classic \cite{draper}.

For such purpose, the organization starts by listing the set $\mathcal{M}$ of
relevant models, which will contain a subset of or all of the models considered above. Then, they must elicit a prior
distribution $p_{D}(M),\, \forall M \in \mathcal{M}$. The defender then performs the ARA analysis on each of those models to obtain their expected utilities
$\psi_{D}(d_{1}|M), \, \forall M \in \mathcal{M}.$ Their expected utility 
taking into account model uncertainty is then given by 
\begin{equation} \label{sim25}
\psi_{D}(d_{1}) = \sum_{M \in \mathcal{M}} p_{D}(M) \psi_{D}(d_{1}|M). 
\end{equation}
Then, their maximum utility decision is $d_{1}^{*} = \arg\max_{d_{1}\in \mathcal{D}_{1}}  \psi_{D}(d_{1}).$ 

Suppose now that the defender is not certain if the malicious insider will be able to act on their own or whether his actions will be affected by other employees. She considers two models: $M_{1},$ the model in Section \ref{ex_simple} and $M_{2},$ that in Section \ref{ex_real}; and elicits a prior probability $p_{D}(M_{1}) = 0.3$.
This implies that $p_{D}(M_{2}) = 0.7$. She would then perform the ARA analysis within each model and arrive at her expected utilities $\psi_{D}(d_{1}|M_{1})$ $= (8.8,-10.65,-48.22),$ for the three options when $p_{D}(a| d_{1})$ is elicited by modeling the attacker's strategic thinking, as illustrated in Section \ref{ex_simple}. Similarly, she  arrives at her expected utilities $\psi_{D}(d_{1}|M_{2})$ $= (-7.21,14.36,-107.6),$ as illustrated in Section \ref{ex_real}. Then, using Eq.~(\ref{sim25}), she can compute her expected utilities $\psi_{D}(d_{1})$ taking into account her model uncertainty $\psi_{D}(d_{1}= (-2.4,6.86,-89.78)$, resulting in {\em investing in Information security} and {\em training of the staff} being the optimal strategy for the defender when taking into account model uncertainty.

\section{Discussion and Further Work} \label{discuss}

 Insider threats constitute a major security problem worldwide. We have develped ARA based models to determine optimal strategies to counter insider threats and illustrated them using a data security application. The olutions proposed in this paper are desirable for the following reasons.
 \begin{itemize}
     \item {They are based on defend-attack-defend models which take into account the existing defensive mechanisms already in place.}
    \item{They take into account whether the malicious insider has been detected or not since this has consequences for the persistence of attacks.}
     \item{In addition, the model in Section \ref{realistic} also models the culture in the organisation that can account for:
     \begin{itemize}
         \item{the security culture in the organisation and its ability to prevent an attack,}
         \item{the security culture to prevent/enable inadvertent attacks, and}
         \item{the inadvertent negative consequences of a not so good culture.}
     \end{itemize}
     }
      \item{They only need the information that a defender is likely to have and do not make unrealistic assumptions such as that of common knowledge or common priors.}
     \item{They are practicably easy to solve and are applicable to any general insider threat problem and not specific to a single problem.}
 \end{itemize}

In general, as in with almost any security application, interactions between the defender and the attacker will expand over several time periods and they will respectively evolve their defenses  and attacks so as to effectively counter their adversarial actions.  This can be modeled using a Markov decision process (MDP). However, a general ARA solution to MDPs has not been developed yet, thus being a promising area for further research. We could then provide a specific MDP solution to the insider threat problem.
This approach could also provide an ARA solution to support the advanced persistent threat (APT) problem, being a persistent and long term threat.

Insider threats come in many different forms. The problem considered here is probably the most obvious, two player version where the malicious insider seeks to harm the organization directly. However, there are more complicated three player versions. These 
could consist of two attackers and one defender or the other way around or even an attacker, a defender and a victim (the victim being a third party). For example, a three player case consisting of a malicious insider, the APT and the organization consists of two attackers and a defender. But the malicious insider could be also be someone who uses their privileges to exploit, abuse or harm a third party. This third party could be clients, customers, students, patients, etc. Therefore, an important extension would be to develop ARA solutions to such complex three player games. This could provide a much more realistic alternative to the game theoretic models proposed in \cite{feng15} and \cite{hu15}.

Players are not always entirely rational and hence incorporating bounded rationality may make the model more realistic. ARA is naturally equipped to incorporate attackers with different reasonings, such as non-strategic thinking, Nash equilibrium, level-$k$ thinking and the mirror equilibrium (\cite{banks16}). However, a general ARA solution using the bounded rationality has not yet been developed. Developing such a solution will enable a bounded rationality ARA solution to the insider threat problem.

ARA relies on the elicitation of the adversary’s utilities and probabilities. Robustness analysis of ARA to these elicitations is necessary, but has yet to be developed.
\cite{rios16} highlight the need and illustrate how a robustness analysis can be performed in principle for ARA. It is important to be able to investigate the  sensitivity of the ARA
outcome, the optimal strategy, to any errors or mis-specifications in the utilities and the probabilities elicited for the analysis.



\section*{Acknowledgment}
The work of CJ was supported by the Strategic Investment funding provided by the University of Waikato. The work of DRI is supported by the AXA-ICMAT Chair on Adversarial Risk Analysis, the Spanish Ministry of Economy and Innovation program MTM2017-86875-C3-1-R and project MTM2015-72907-EXP. Work supported by the EU's Horizon 2020 project 740920 CYBECO (Supporting Cyberinsurance from a Behavioural Choice Perspective) as well as 
as well as the US National
Science Foundation through grant DMS-163851.

\bibliography{ARA_insider}

\begin{thebibliography}{}

\bibitem[Antos and Pfeffer, 2010]{Antos10}
Antos, D. and Pfeffer, A. (2010).
\newblock Representing bayesian games without a common prior.
\newblock In {\em Proceedings of the 9th International Conference on Autonomous
  Agents and Multiagent Systems: Volume 1 - Volume 1}, AAMAS '10, pages
  1457--1458.

\bibitem[{Axelrad} et~al., 2013]{Axelrad13}
{Axelrad}, E.~T., {Sticha}, P.~J., {Brdiczka}, O., and {Shen}, J. (2013).
\newblock A bayesian network model for predicting insider threats.
\newblock In {\em 2013 IEEE Security and Privacy Workshops}, pages 82--89.

\bibitem[Banks et~al., 2015]{banks16}
Banks, D., Rios, J., and Insua, D.~R. (2015).
\newblock {\em Adversarial Risk Analysis}.
\newblock CRC Press, first edition.

\bibitem[Brdiczka et~al., 2012]{oliver12}
Brdiczka, O., Liu, J., Shen, J., Patil, A., Chow, R., Bart, E., and Ducheneaut,
  N. (2012).
\newblock Proactive insider threat detection through graph learning and
  psychological context.
\newblock {\em IEEE CS Security and Privacy Workshop}.

\bibitem[Brown et~al., 2006]{BROWN}
Brown, G., Carlyle, M., Salmeron, J., and Wood, R. (2006).
\newblock Defending critical infrastructure.
\newblock {\em Interfaces}, 36:530--544.

\bibitem[Brown and Cox, 2011]{brown11}
Brown, G.~G. and Cox, Jr., L.~A. (2011).
\newblock Making terrorism risk analysis less harmful and more useful: Another
  try.
\newblock {\em Risk Analysis}, 31(2):193--195.

\bibitem[Camerer, 2003]{camerer03}
Camerer, C. (2003).
\newblock {\em Behavioural Game Theory}.
\newblock Princeton University Press.

\bibitem[CERT, 2012]{cert2012}
CERT (2012).
\newblock {\em 2012 Cyber Security Watch Survey. How Bad is the Insider
  Threat?}
\newblock Software Engineering Institute, Carnegie Mellon.

\bibitem[Choi et~al., 2018]{Choi18}
Choi, S., Martins, J.~T., and Bernik, I. (2018).
\newblock Information security: Listening to the perspective of organisational
  insiders.
\newblock {\em Journal of Information Science}, 44(6):752--767.

\bibitem[Colwill, 2009]{Colwill09}
Colwill, C. (2009).
\newblock Human factors in information security: The insider threat – who can
  you trust these days?
\newblock {\em Information Security Technical Report}, 14(4):186 -- 196.

\bibitem[Cox, 2009]{cox09a}
Cox, Jr., L.~A. (2009).
\newblock Game theory and risk analysis.
\newblock {\em Risk Analysis}, 29(8):1062--1068.

\bibitem[Draper, 1995]{draper}
Draper, D. (1995).
\newblock Assessment and propagation of model uncertainty.
\newblock {\em Journal Royal Statistical Society}, 57(1):45 -- 97.

\bibitem[Esteban and Insua, 2014]{esteban14}
Esteban, P.~G. and Insua, D.~R. (2014).
\newblock Supporting an autonomous social agent within a competitive
  environment.
\newblock {\em Cybernetics and Systems}, 45(3):241--253.

\bibitem[Ezell et~al., 2010]{ezell10}
Ezell, B., Bennett, S., Winterfeldt, D., Sokolowski, J., and Collins, A.
  (2010).
\newblock Probabilistic risk analysis and terrorism risk.
\newblock {\em Risk Analysis}, 30(4).

\bibitem[Feng et~al., 2015]{feng15}
Feng, X., Zheng, Z., Hu, P., Cansever, D., and Mohapatra, P. (2015).
\newblock Stealthy attacks meets insider threats: A three-player game model.
\newblock In {\em MILCOM 2015 - 2015 IEEE Military Communications Conference},
  pages 25--30.

\bibitem[Gil and Parra-Arnau, 2019]{Gil19}
Gil, C. and Parra-Arnau, J. (2019).
\newblock An adversarial-risk-analysis approach to counterterrorist online
  surveillance.
\newblock {\em Sensors}, 19(3).

\bibitem[Gil et~al., 2016]{gil16}
Gil, C., Rios~Insua, D., and Rios, J. (2016).
\newblock Adversarial risk analysis for urban security resource allocation.
\newblock {\em Risk Analysis}, 36(4):727--741.

\bibitem[Gintis, 2009]{gintis09}
Gintis, H. (2009).
\newblock {\em The Bounds of Reason: Game Theory and the Unification of
  Behavioural Sciences.}
\newblock Princeton University Press.

\bibitem[Greitzer et~al., 2012]{GREIZER}
Greitzer, F., Dalton, A., Kangas, L., Noonan, C., and Hohimer, R. (2012).
\newblock Identifying at-risk employees: Modeling psychosocial precursors of
  potential insider threats.
\newblock {\em Proc. 25th HICSS}.

\bibitem[Harsanyi, 1967]{harsanyi67}
Harsanyi, J. (1967).
\newblock Games with incomplete information played by "bayesian" players,
  i-iii.
\newblock {\em Management Science}, 14(3).

\bibitem[Hu et~al., 2015]{hu15}
Hu, P., Li, H., Fu, H., Cansever, D., and Mohapatra, P. (2015).
\newblock Dynamic defense strategy against advanced persistent threat with
  insiders.
\newblock In {\em 2015 IEEE Conference on Computer Communications (INFOCOM)},
  pages 747--755.

\bibitem[Hunker and Probst, 2009]{HUNKER}
Hunker, J. and Probst, C. (2009).
\newblock Insiders and insider threats: An overview of definitions and
  mitigation techniques.
\newblock {\em Journal Wireless Mobile Networks, Ubiquitous Computing, and
  Dependable Applications}, 2(1):4--27.

\bibitem[Insua et~al., 2009]{rios09}
Insua, I.~R., Rios, J., and Banks, D. (2009).
\newblock Adversarial risk analysis.
\newblock {\em Journal of the American Statistical Association},
  104(486):841--854.

\bibitem[Kantzavelou and Katsikas, 2010]{kant10}
Kantzavelou, I. and Katsikas, S. (2010).
\newblock A game-based intrusion detection mechanism to confront internal
  attackers.
\newblock {\em Computers \& Security}, 29(8):859 -- 874.

\bibitem[Lee and Rotoloni, 2015]{DAT2015}
Lee, W. and Rotoloni, B. (2015).
\newblock {\em Emerging Cyber Threat Report 2015}.
\newblock Georgia Tech Information Security Centre and Georgia Tech Research
  Institute.

\bibitem[Liu et~al., 2008]{liu08}
Liu, D., Wang, X., and Camp, J. (2008).
\newblock Game-theoretic modeling and analysis of insider threats.
\newblock {\em International Journal of Critical Infrastructure Protection},
  I:75 -- 80.

\bibitem[Liu et~al., 2009]{liu*}
Liu, D., Wang, X., and Camp, J. (2009).
\newblock Mitigating inadvertent insider threats with incentives.
\newblock pages 1--16.

\bibitem[Martinez-Moyano et~al., 2008]{martinez08}
Martinez-Moyano, I., Rich, E., Conrad, S., Andersen, D., and Stewart, T.
  (2008).
\newblock A behavioral theory of insider-threat risks: a system dynamic
  approach.
\newblock {\em ACM Transactions on Modeling and Computer Simulation}, 18(2).

\bibitem[Moore et~al., 2015]{moore15}
Moore, A., Novak, W., Collins, M., Trzeciak, R., and Theis, M. (2015).
\newblock {\em Effective Insider Threat Programs: Understanding and Avoiding
  Potential Pitfalls}.
\newblock White Paper.

\bibitem[Naveiro et~al., 2019]{naviero19}
Naveiro, R., Redondo, A., Insua, D.~R., and Ruggeri, F. (2019).
\newblock Adversarial classification: An adversarial risk analysis approach.
\newblock {\em International Journal of Approximate Reasoning}, 113:133 -- 148.

\bibitem[Probst et~al., 2010]{Probst10}
Probst, C., Hunker, J., Bishop, M., and Gollman, D. (2010).
\newblock {\em Insider Threats in Cyber Security}.
\newblock Springer US.

\bibitem[Raiffa et~al., 2002]{raiffa02}
Raiffa, H., Richardson, J., and Metcalfe, D. (2002).
\newblock {\em Negotiation Analysis}.
\newblock Havard University Press.

\bibitem[Rios and Insua, 2012]{rios12}
Rios, J. and Insua, D.~R. (2012).
\newblock Adversarial risk analysis for counterterrorism modeling.
\newblock {\em Risk Analysis}, 32(5):894--915.

\bibitem[R{\'i}os~Insua et~al., 2019]{cookebook}
R{\'i}os~Insua, D., Banks, D., Ríos, J., and Ortega, J. (2019).
\newblock {\em Adversarial Risk Analysis as an Expert Judgement Methodology},
  pages~--.
\newblock Springer International Publishing.

\bibitem[Rios~Insua et~al., 2019]{rios19}
Rios~Insua, D., Couce-Vieira, A., Rubio, J.~A., Pieters, W., Labunets, K., and
  G.~Rasines, D. (2019).
\newblock An adversarial risk analysis framework for cybersecurity.
\newblock {\em Risk Analysis}, n/a(n/a).

\bibitem[R{\'i}os~Insua et~al., 2016]{rios16}
R{\'i}os~Insua, D., Ruggeri, F., Alfaro, C., and Gomez, J. (2016).
\newblock {\em Robustness for Adversarial Risk Analysis}, pages 39--58.
\newblock Springer International Publishing.

\bibitem[Safa, 2017]{Safa17}
Safa, N. (2017).
\newblock The information security landscape in the supply chain.
\newblock {\em Computer Fraud and Security}, 2017(6):16--20.

\bibitem[S{\'a}kovics, 2001]{Sakovics01}
S{\'a}kovics, J. (2001).
\newblock Games of incomplete information without common knowledge priors.
\newblock {\em Theory and Decision}, 50(4):347--366.

\bibitem[Sarkar, 2010]{SARKAR}
Sarkar, K. (2010).
\newblock Assessing insider threats to information security using technical,
  behavioural and organisational measures.
\newblock {\em Info. Sec. Tech. Rep.}, 15:112--133.

\bibitem[Schulze, 2018]{ca2018}
Schulze, H. (2018).
\newblock {\em Insider Threat, 2018 report}.
\newblock ca Technologies.

\bibitem[Sevillano et~al., 2012]{sevillano12}
Sevillano, J.~C., Insua, D.~R., and Rios, J. (2012).
\newblock {Adversarial Risk Analysis: The Somali Pirates Case}.
\newblock {\em Decision Analysis}, 9(2):86--95.

\bibitem[Silowash et~al., 2012]{SILOWASH}
Silowash, G., Cappelli, D., Moore, A., Trzeciak, R., Shimeall, T., and Flynn,
  L. (2012).
\newblock Common sense guide to mitigating insider threats.
\newblock {\em Def. Tech. Inf. Center Tech. Report}.

\bibitem[Tang et~al., 2011]{tang11}
Tang, K., Zhao, M., and Zhou, M. (2011).
\newblock Cyber insider threats situation awareness using game theory and
  information fusion-based user behavior predicting algorithm.
\newblock {\em Journal of Information \& Computational Science}, 8(3):529 --
  545.

\bibitem[Wang and Bier, 2013]{wang13}
Wang, C. and Bier, V. (2013).
\newblock Expert elicitation of adversary preferences using ordinal judgments.
\newblock {\em Operations Research}, 61(2):372--385.

\bibitem[Wang and Banks, 2011]{wang11}
Wang, S. and Banks, D. (2011).
\newblock Network routing for insurgency: An adversarial risk analysis
  framework.
\newblock {\em Naval Research Logistics (NRL)}, 58(6):595--607.

\bibitem[Ware, 2017]{haystax2017}
Ware, B. (2017).
\newblock {\em Insider Attacks, 2017 insider threat study}.
\newblock Haystax.

\bibitem[Wood et~al., 2016]{sym2016}
Wood, P., Nahorney, B., Chandrasekar, K., Wallace, S., and Haley, K. (2016).
\newblock {\em Internet Security Threat Report}, volume~21.
\newblock Symantec.

\end{thebibliography}
\bibliographystyle{apalike}
\end{document}